\documentstyle[epsfig]{mn}

\def\simgt{\mathrel{\lower0.6ex\hbox{$\buildrel {\textstyle >}
 \over {\scriptstyle \sim}$}}}
\def\simlt{\mathrel{\lower0.6ex\hbox{$\buildrel {\textstyle <}
 \over {\scriptstyle \sim}$}}}

\newcommand{\Msolar}{\mbox{\,$\rm M_{\odot}$}}        

\newcommand{\mg}{Mg{\sc ii}\,\,}
\newcommand{\hb}{H$\beta$\,\,}
\title[Measuring the black hole masses of high redshift quasars]
{Measuring the black hole masses of high redshift quasars}
\author[R.J. McLure and M.J. Jarvis]
{R.J. McLure$^1$ and M.J. Jarvis$^2$\\
$^{1}$Institute for Astronomy, University of Edinburgh, Royal
Observatory, Edinburgh EH9 3HJ.\\
$^{2}$Sterrewacht Leiden, Postbus 9513, 2300 RA Leiden, the Netherlands\\}
\date{Submitted for publication in MNRAS}
\begin{document}
\maketitle
\begin{abstract}
A new technique is presented for determining the black-hole masses of 
high-redshift quasars from optical spectroscopy. The new method utilizes 
the full-width half maximum (FWHM) of the low-ionization Mg{\sc ii} 
emission line and 
the correlation between broad-line region (BLR) radius and 
continuum luminosity at 3000\AA. 
Using archival UV spectra it is found that the 
correlation between BLR radius and 3000\AA\, luminosity is tighter 
than the established correlation with 5100\AA\, luminosity. 
Furthermore, it is found that the correlation between BLR radius and 
3000\AA\, continuum luminosity is consistent 
with a relation of the form $R_{BLR}\propto \lambda
L_{\lambda}^{0.5}$, as expected for a constant ionization parameter. 
Using a sample of objects with broad-line radii determined from 
reverberation mapping it is shown that the FWHM of Mg {\sc ii} and 
H$\beta$ are consistent with following an exact one-to-one 
relation, as 
expected if both H$\beta$ and Mg{\sc ii} are emitted at the same
radius from the central ionizing source. 
The resulting virial black-hole mass estimator based on 
rest-frame UV observables is shown to reproduce black-hole mass 
measurements based on reverberation mapping to within 
a factor of 2.5 ($1\sigma$). Finally, the new UV 
black-hole mass estimator is shown to produce identical results to 
the established optical (H$\beta$) estimator when applied to 128
intermediate-redshift ($0.3<z<0.9$) quasars drawn from 
the Large Bright Quasar Survey and the radio-selected Molonglo quasar 
sample. We therefore conclude that the new UV virial black-hole mass 
estimator can be reliably used to estimate the black-hole masses of 
quasars from $z\sim0.25$ through to the peak epoch of quasar activity 
at $z\sim2.5$ via optical spectroscopy alone.
\end{abstract}

\section{Introduction}
\label{intro}
A reliable method for estimating the black-hole masses of 
high-redshift quasars would provide crucial new information for 
understanding the nature and cosmological evolution of
quasars. In this paper we address this problem by deriving a virial
black-hole mass estimator based on rest-frame UV observables which
is capable of producing reliable quasar black-hole mass estimates 
from optical spectra out to redshifts of $z\sim2.5$.

Currently, the most direct measurements of the
central black-hole masses of powerful active galactic nuclei (AGN) 
come from the reverberation mapping studies of 17 Seyferts and 
17 PG quasars by Wandel, Peterson \& Malkan (1999) and Kaspi et
al. (2000) respectively. By measuring the time-lag
between the variations of the AGN continuum and the broad \hb (4861
\AA) emission line, these long-term monitoring
programmes have delivered direct measurements of the radius of the \hb
emission region from the central ionizing source 
($R_{BLR}=c\tau$, where $\tau$ is the time lag between the continuum
and \hb variations). By combining 
these measurements with the assumption that the FWHM of 
the \hb line reflects the virialized bulk motion of the 
line-emitting material, it is then possible to produce the 
so-called virial estimate of the central black 
hole mass (see Wandel, Peterson \& Malkan 1999 for a full discussion). 
Although the virial mass estimate is potentially subject to 
many uncertainties (eg. Krolik 2001), comparison with the predictions
of the completely independent relation between black-hole mass and
host-galaxy stellar velocity dispersion have shown the two to be in 
excellent agreement (Ferrarese et al. 2001, Gebhardt et al. 2000). 

Unfortunately, because of the many years of monitoring required, it is
unrealistic to expect that reverberation mapping measurements can be
obtained for large samples of quasars, even at low redshift. However, 
the existing reverberation mapping of low redshift AGN has 
revealed a correlation between $R_{BLR}$ and the monochromatic 
AGN continuum luminosity at 5100\AA\, 
(eg. $R_{BLR}\propto \lambda L_{5100}^{0.7}$; Kaspi et al. 2000). 
By exploiting this correlation to estimate $R_{BLR}$ it is 
possible to produce a virial black-hole mass estimate from a 
single spectrum covering the \hb emission line.

This technique has recently been widely employed to investigate how the
masses of quasar black holes relate to the
properties of the surrounding host galaxies (eg. McLure \& 
Dunlop 2002;2001; Laor 2001) and the radio luminosity of the 
central engine (Dunlop et al. 2002; Lacy et al. 2001). However, 
because of the atmospheric restrictions and 
relative observational expense of near-infrared spectroscopy, 
the vast majority of previous studies 
have relied on optical spectroscopy of the \hb region. Unfortunately, 
this effectively enforces a redshift upper limit of $z\simeq 0.8-0.9$,
with the vast majority of studies concentrating on samples at $z\leq0.3$. 

An ideal solution to this problem would be to
calibrate a rest-frame UV emission line and continuum measurement such
that it becomes possible to produce a virial black-hole mass estimate for 
high-redshift quasars from relatively straightforward 
optical spectroscopy. It is
the purpose of this paper to provide the required UV calibration. There are 
two strong, permitted, broad UV lines which are candidates to replace
\hb in the virial black-hole mass estimate;
C{\sc iv}(1549\AA) and Mg {\sc ii}(2798\AA). We note here that 
Vestergaard (2002) recently derived a UV virial black-hole
mass estimator using the C{\sc iv} FWHM and the continuum luminosity
at 1350\AA\,. Although C{\sc iv} does have the advantage of being 
accessible with optical spectroscopy at the very highest redshifts
($3<z<5$), where Mg{\sc ii} is redshifted into the near-infrared, there 
are a number of reasons to believe that Mg{\sc ii} represents a 
better substitute for H$\beta$ over the redshift range $0.8<z<2.5$.

The main reason for adopting \mg as the UV tracer of BLR velocity is
that, like H$\beta$, Mg{\sc ii} is a low-ionization line. Furthermore,
due to the similarity of their ionization potentials, it is reasonable 
to expect that the Mg{\sc ii} and H$\beta$ emission lines are 
produced by gas at virtually the same radius from the central
ionizing source, an assumption which is supported by
the good agreement between the Mg{\sc ii} and H$\beta$ FWHM (see
Section 5). This has two important consequences. Firstly, it
allows us to directly adopt the reverberation mapping determinations
of $R_{BLR}$ when calibrating the correlation between $R_{BLR}$ and
3000\AA\, continuum luminosity. Secondly, because the
line-widths of \mg and \hb should trace the same BLR velocities, we are
able to simply substitute the FWHM of \mg for that of \hb in the
virial mass estimator. 

There is one further practical advantage which favours \mg over C{\sc iv} 
as the UV substitute for \hb. Due to the fact that \mg becomes
accessible to optical spectroscopy at $z\sim0.3$, both \hb and \mg can
be observed in a single optical spectrum within the redshift interval 
$0.3 \leq z \leq 0.9$. Consequently, it is therefore possible to 
directly compare the results of the new UV virial mass 
estimator against those of the
established optical virial mass estimator for objects at intermediate
redshifts. This test is performed in Section \ref{lbqs} using data for
the Large Bright Quasar Survey (Forster et al. 2001) and the 
radio-selected Molonglo quasar sample (Baker et al. 1999).

The structure of this paper can be summarized as follows. 
In Section \ref{virial} we
briefly review the main aspects of the virial black-hole mass estimate, 
before proceeding in Section \ref{samples} to describe the properties 
of the sample of low-redshift AGN with reverberation mapping results.
In Sections \ref{radius} and \ref{proxy} this sample is used to 
calibrate the relation between
broad-line radius and 3000\AA\, continuum luminosity 
($R_{BLR}-\lambda L_{3000}$) and to demonstrate the viability of 
adopting \mg as a direct substitute for \hb in the virial 
black-hole mass estimator. Our final calibration of the UV 
black-hole mass estimator is presented in Section \ref{uv}. In 
Section \ref{lbqs} the effectiveness of the new UV black-hole 
mass estimator is 
demonstrated by application to members of the Large Bright Quasar
Survey (LBQS) and the Molonglo quasar sample (MQS) for which
 both \hb and \mg FWHM measurements are available. Our conclusions are
presented in Section \ref{conc}. All cosmological calculations 
presented in this paper assume $\Omega_{m}=0.3, \Lambda=0.7,
H_{0}=70$ kms$^{-1}$Mpc$^{-1}$.

\section{The virial black-hole mass estimate}
\label{virial}
The underlying assumption behind the method of estimating AGN black-hole
masses from the width of their broad emission lines is that the motion of the
line-emitting material is virialized. Under this assumption the width
of the broad lines can be used to trace the Keplerian velocity 
of the broad-line gas, and thereby allow an estimate of the 
central black-hole mass via the formula :
\begin{equation}
M_{bh}=G^{-1}R_{BLR}V_{BLR}^{2}
\label{mass}
\end{equation}
\noindent
where $R_{BLR}$ is the BLR radius and $V_{BLR}$ is the
Keplerian velocity of the BLR gas. Recent evidence to support the
Keplerian interpretation of AGN broad-line widths has been 
presented by Peterson \& Wandel (2000) and Onken \& Peterson (2002). These
authors demonstrate, at least for several Seyfert galaxies which
have been well studied for reverberation mapping purposes, that the
FWHM of various emission lines follow the 
$V_{BLR}\propto R_{BLR}^{-0.5}$ relation expected for Keplerian motion. 
It is important to note that it is not clear that the virial condition 
is satisfied for the high-ionization C{\sc iv} emission line. Substantial
evidence exists in the literature that 
C{\sc iv} may be produced in some form of 
outflow, as is suggested by the systematic blue-shift
of C{\sc iv} compared to H$\beta$ (eg. Marziani et al. 1996).

\subsection{Broad-line region geometry}
In Eqn \ref{mass} the velocity of the BLR gas is 
taken as $V_{BLR}=$~$f\times$~$ H\beta$(FWHM),
where $f$ is a geometric factor which relates the \hb FWHM to 
the intrinsic Keplerian velocity. Due to the fact that there 
is currently no consensus on the 
geometry of the BLR in radio-quiet quasars it is conventional to set
$f=\sqrt{3}/2$. However, it is clear that if AGN broad lines are
in general produced in a more disc-like configuration, evidence for which has
been found in radio-loud quasars by numerous authors (eg. Wills \&
Browne 1986; Brotherton 1996; Vestergaard, Wilkes \&\ Barthel 2000),
then $f$ will be inclination dependent. Indeed, 
in McLure \& Dunlop (2002) evidence was presented that the \hb FWHM 
is inclination dependent in radio-quiet quasars, and more consistent
with the orbits of the \hb emitting material having a flattened 
disc-like geometry than being randomly orientated. However, even in the
scenario where the FWHM of \hb and \mg are entirely orientation
dependent, inclination will only have a significant effect on the virial
black-hole mass estimate for those objects within $\sim 15^{\circ}$ of the
line of sight. If quasars are taken to be distributed randomly within
$\sim 45^{\circ}$ of the line of sight (Barthel 1989) 
then this will only effect some
10\% of objects. Consequently, for the purposes of providing a 
generally applicable UV black-hole mass estimator we 
adopt $f=1$ throughout this paper. 
\section{The reverberation mapped sample}
\label{samples}
\begin{table*}
\begin{tabular}{lcccccccc}
\hline
Object & z & $\log\left(\lambda L_{3000}\right)$  &
$\log \left( \lambda L_{5100} \right) $  & $\log \left( R_{BLR} \right)$&
Mg{\sc ii} FWHM&&H$\beta$ FWHM& \\
          &       & /W         & /W         & /lt-days &/kms$^{-1}$ &
&/kms$^{-1}$ \\          
\hline
PG 0026+129&  0.142&38.23&       38.07&       2.05&\phantom{0000}& &\phantom{0}1860&1\\      
PG 0052+251&  0.155&38.29&       38.04&       2.13&\phantom{0000}& &\phantom{0}5200&1\\      
PG 0804+761&  0.100&38.40&       37.96&       2.19&\phantom{0000}& &\phantom{0}3070&1\\        
PG 0844+349&  0.064&37.61&       37.44&       1.38&3800&2&\phantom{0}3800&1\\           
PG 0953+414&  0.239&38.69&       38.41&       2.18&\phantom{0000}&&\phantom{0}3130&1\\        
PG 1211+143&  0.085&38.20&       38.00&       2.00&2050&5&\phantom{0}1860&1\\           
PG 1226+023&  0.158&39.23&       38.97&       2.59&3160&5&\phantom{0}3520&1\\          
PG 1229+204&  0.064&37.53&       37.33&       1.70&2560&5&\phantom{0}3360&1\\           
PG 1307+085&  0.155&38.26&       38.01&       2.09&\phantom{0000}&&\phantom{0}5320&1\\        
PG 1351+640&  0.087&37.93&       37.79&       2.36&2790&5&\phantom{0}5660&1\\           
PG 1411+442&  0.089&37.68&       37.53&       2.01&1940&5&\phantom{0}2670&1\\           
PG 1426+015&  0.086&38.11&       37.78&       1.98&\phantom{0000}&&\phantom{0}6820&1\\        
PG 1613+658&  0.129&37.85&       37.74&       1.59&\phantom{0000}&&\phantom{0}8450&1\\        
PG 1617+175&  0.114&38.16&       37.74&       1.93&\phantom{0000}&&\phantom{0}5330&1\\        
PG 1700+518&  0.292&38.75&       38.70&       1.94&\phantom{0000}&&\phantom{0}2210&1\\        
PG 1704+608&  0.371&38.84&       38.72&       2.50&\phantom{0000}&&\phantom{0}6560&1\\          
PG 2130+099&  0.061&37.77&       37.47&       2.30&4360&2&\phantom{0}2330&1\\           
3C 120     &  0.035&37.80    &   36.94    &   1.62&3360&2&\phantom{0}1910&3    \\      
3C 390     &  0.052&36.49    &   36.89    &   1.36&8410&2&10000&3\\  
Akn 120    &  0.035&37.60    &   37.22    &   1.57&9660&2&\phantom{0}5800&3     \\    
F9         &  0.047&37.62    &   37.22    &   1.21&5680&2&\phantom{0}6080&4     \\     
IC 4329A   &  0.016&34.25    &   36.28    &   0.15&\phantom{0000}&&\phantom{0}5050&3   \\  
Mrk 79     &  0.022&36.42    &   36.70    &   1.26&4730&2&\phantom{0}4470&3  \\     
Mrk 110    &  0.035&36.77    &   36.66    &   1.27&2420&2&\phantom{0}2120&1     \\    
Mrk 335    &  0.026&37.15    &   36.87    &   1.21&8605&2&\phantom{0}5685&6     \\    
Mrk 509    &  0.033&37.67    &   37.24    &   1.88&4790&2&\phantom{0}2270&4    \\     
Mrk 590    &  0.026&36.88    &   36.78    &   1.30&4370&2&\phantom{0}2470&3     \\    
Mrk 817    &  0.032&36.99    &   36.80    &   1.19&5770&2&\phantom{0}4490&3 \\   
NGC 3783   &  0.009&36.32    &   36.31    &   0.65&2690&2&\phantom{0}3780&4       \\   
NGC 3227   &  0.004&35.18    &   35.37    &   1.04&\phantom{0000}&&\phantom{0}4920&3 \\    
NGC 4051   &  0.003&35.08    &   34.78    &   0.77&2790&2&\phantom{0}1170&3     \\  
NGC 4151   &  0.003&34.93    &   35.92    &   0.48&3990&2&\phantom{0}5910&3      \\ 
NGC 5548   &  0.014&36.52    &   36.50    &   1.33&5420&2&\phantom{0}6200&4    \\    
NGC 7469   &  0.017&36.95    &   36.81    &   0.69&3258&2&\phantom{0}3000&3      \\   
\hline
\end{tabular}
\caption{Data for the reverberation mapped sample. Columns 1 and 2
detail the name and redshift of each object. Columns 3 and 4 list the
monochromatic luminosities at 3000\AA\, and 5100\AA\,
respectively. The 3000\AA\, and 5100\AA\, luminosities for the PG quasars
are derived from the spectrophotometric study of 
Neugebauer et al. (1987). For the Seyfert galaxies the
3000\AA\, luminosities are derived from new fits to IUE archive spectra,
while the 5100\AA\, luminosities are taken from Kaspi et al. (2000)
after conversion to our cosmology. Column 5 lists the BLR 
radius measurements from Kaspi et al. (2000), expect for
NGC 4051 which is taken from Peterson et al. (2000). Columns 6 \& 7 
detail the FWHM of the Mg{\sc ii} and H$\beta$ emission 
lines respectively. The references for
the FWHM measurements are as follows : 1. Boroson \& Green (1992),
2. This work, 3. Peterson, Wandel \& Malkan (1999), 
4. Marziani et al. (1996), 5. Corbin \& Boroson (1996) and 6. Zheng et
al. (1995)} 
\label{maintab}
\end{table*}
As mentioned previously, the sample of 17 Seyferts and 17
PG quasars with reverberation mapping measurements (hereafter the RM
sample) currently represents the best available set of 
black-hole mass estimates of powerful broad-line AGN. Therefore, it is
the objects from this sample which we will use to calibrate the new UV
virial black-hole mass estimator. Optical and UV data for the 
combined 34-object RM sample are listed in Table \ref{maintab}. 
The $R_{BLR}$ data are taken from Kaspi et al. (2000), with the 
exception of NGC 4051 for
which we take the updated figure of $R_{BLR}=5.9$ light-days 
from Peterson et al. (2000). 
\subsection{Optical data}
The 5100\AA\, luminosities for the PG quasars have been calculated
using the fluxes from the spectrophotometric study of 
Neugebauer et al. (1987). 
For the Seyfert galaxies the 5100\AA\, luminosities 
are from Kaspi et al. (2000) after conversion to our cosmology. The
\hb FWHM values for the PG quasars are taken from Boroson \& Green
(1992). The \hb FWHM values for the Seyfert galaxies are drawn from a
number of literature sources which are listed in the 
caption to Table \ref{maintab}.
\subsection{UV data}
As with the optical continuum luminosities, the 3000\AA\, luminosities
listed in 
Table \ref{maintab} for the PG quasars are derived 
from Neugebauer et al. (1987). 
In contrast, the 3000\AA\, luminosities for the Seyfert galaxies,
together with 17/22 of the \mg FWHM values, are new measurements based
on our fitting of spectra from the International Ultraviolet
Explorer (IUE) final data archive. A full discussion of our
line-fitting method will be presented in a follow-up paper (McLure et
al. 2002, in prep). In brief, we fit the region spanning the \mg
emission line (2300\AA $\rightarrow$ 3100\AA) with a combination of
a power-law continuum, a Fe{\sc ii} emission template 
and two gaussians representing the broad and narrow 
components of the \mg emission line.

As with the H$\beta$ emission line, it is crucial to account for
blending with broad Fe{\sc ii} emission in order to obtain a reliable
measurement of the Mg{\sc ii} line-width. Consequently, following the
analysis of Corbin \& Boroson (1996) we have constructed a template of
the FeII emission around the Mg{\sc ii} line from an archival IUE
spectrum of the narrow line Seyfert galaxy I~Zw~1. During the fitting
process the amplitude and FWHM of the FeII template were left as free
parameters in order to obtain the best possible match to the FeII
emission of each object. 

\section{Estimating the BLR radius}
\label{radius}
\begin{figure*}
\centerline{\epsfig{file=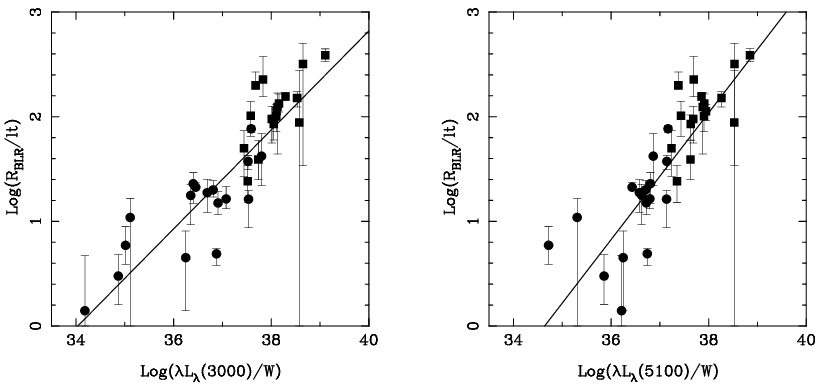,width=14cm,angle=0}}
\caption{The correlation between broad-line radius and AGN continuum
luminosity at 3000\AA\, and 5100\AA\ . The 17 PG quasars are shown as
filled squares and the 17 Seyfert galaxies are shown as filled circles.
The best-fitting BCES bisector fits to both correlations are shown as
solid lines, and correspond to $R_{BLR}\propto \lambda
L_{3000}^{0.47}$ and $R_{BLR}\propto \lambda L_{5100}^{0.61}$
respectively.}
\label{rl}
\end{figure*}
In order to provide an estimate of $R_{BLR}$ it is necessary, 
in the absence of reverberation mapping, to exploit 
the known correlation between $R_{BLR}$ and AGN continuum luminosity. 
By combining the results of their reverberation mapping observations
with those of Wandel, Peterson \& Malkan (1999), Kaspi et al. (2000) 
investigated the correlation between $R_{BLR}$ and 5100\AA\, 
continuum luminosity and found a best-fit 
of the form $R_{BLR}\propto \lambda L_{5100}^{0.70\pm 0.03}$. In their
re-analysis of the same correlation, Peterson et al. (2000) 
found a best fitting relation of the form $R_{BLR}\propto \lambda
L_{5100}^{0.62\pm 0.02}$. As well as being important within the
context of the virial black-hole mass estimate, the slope of the
$R_{BLR}-\lambda L_{\lambda}$ relation is of intrinsic interest 
because it provides information
about the AGN ionization parameter $U$ ($U=\frac{Q}{4\pi r^{2}c}$,
where $Q=\int \frac{L_{\nu}}{h\nu} d\nu$). For example, neither of 
the two fits mentioned above are consistent with the relation 
$R_{BLR}\propto \lambda L_{\lambda}^{0.5}$ expected if 
the ionization parameter 
is approximately the same for all AGN, regardless of luminosity. 
In this section we investigate the 
slope of the $R_{BLR}-\lambda L_{\lambda}$ relation using continuum 
luminosity measurements at both 3000\AA\, and 5100\AA.

\subsection{Regression analysis}
Given the potential influence of the $R_{BLR}-\lambda L_{\lambda}$ relation
upon the accuracy of the virial black-hole mass estimator, we have chosen to
employ three different regression techniques in our analysis to
explore the possible variation in the best-fitting slope. The first 
technique is a straightforward weighted least-squares 
fit (WLS), in which the fit is weighted by the error in the
reverberation mapping radius (Press et al. 1992). The second
technique is a robust iterative chi-square fit (FITxy) in which the errors in
both variables are taken into account (Press et al. 1992). However, although
FITxy takes into account the errors on both variables, it does not
account for the fact that there is likely to be intrinsic scatter in
the $R_{BLR}-\lambda L_{\lambda}$ relation. To account for this, the third
technique we have employed is the BCES estimator of Akritas \&
Bershady (1996) which accounts for errors in both variables and 
intrinsic scatter. 

\subsection{The radius-luminosity relation at 3000\AA\, and 5100\AA\,}

The results of the fitting process (in $\log-\log$ space) 
of both the $R_{BLR}-\lambda L_{3000}$ and $R_{BLR}-\lambda
L_{5100}$ relations are listed in Table \ref{rltab} where the listed
BCES fit is the bisector. It can be seen from Table \ref{rltab} that
the results of all three regression methods are consistent with each
other for both the $R_{BLR}-\lambda L_{5100}$ and $R_{BLR}-\lambda
L_{3000}$ relations. Furthermore, it can be seen that all three
regression methods indicate that the $R_{BLR}-\lambda
L_{3000}$ relation is fully consistent with a slope of $\lambda
L_{\lambda}^{0.5}$, as expected for a constant ionization parameter.

The three fits to the $R_{BLR}-\lambda L_{5100}$ relation indicate
that a slope of a $\lambda L_{\lambda}^{0.6}$ is more 
appropriate for our choice of continuum data, in good agreement with
the fit of Peterson et al. (2000). This is flatter
than the slope of $\lambda L_{\lambda}^{0.7}$ previously determined for 
the $R_{BLR}-\lambda L_{5100}$ relation by Kaspi et al. (2000),
although the two are formally consistent. 

There are two effects which are responsible for our 
finding of a flatter slope in the $R_{BLR}-\lambda L_{5100}$ 
relation. Firstly, our choice of
cosmology ($\Omega_{m}=$~$0.3, \Lambda=$~$0.7,
H_{0}=70$~kms$^{-1}$Mpc$^{-1}$) 
results in higher luminosities for the 17 PG
quasars in the RM sample, compared to the 
$\Omega_{m}$~$=1$, $\Lambda=$~$0, H_{0}=75$~kms$^{-1}$Mpc$^{-1}$ cosmology
adopted by Kaspi et al., while having little effect on the
luminosities of the lower redshift Seyfert galaxies. 
Secondly, further flattening of the slope is due to the fact that 
the 5100\AA\, fluxes for the PG quasars adopted here, derived from 
the Neugebauer et al. (1987) data, are typically $\sim 0.1$ dex higher 
than those determined by Kaspi et al. Indeed, if we adopt the same PG quasar 
luminosities and cosmology
as Kaspi et al., then the BCES bisector fit to the $R_{BLR}-\lambda
L_{5100}$ relation becomes: $R_{BLR} \propto \lambda
L_{5100}^{0.65 \pm 0.11}$, entirely consistent with the slope
determined by Kaspi et al.

Throughout the remainder of the paper we adopt the BCES bisector fits
as our best estimate of the $R_{BLR}-\lambda L_{\lambda}$ relation at
both 3000\AA\, and 5100\AA\,. The bisector fits to both data-sets are
shown in Fig \ref{rl} and have linear versions as follows:
\begin{equation}
R_{BLR}=(25.2\pm3.0) \left[\lambda L_{3000}/10^{37}W
\right]^{(0.47\pm0.05)}
\end{equation}
\begin{equation}
R_{BLR}=(26.4\pm4.4) \left[\lambda L_{5100}/10^{37}W
\right]^{(0.61\pm0.10)}
\end{equation}
\noindent
where $R_{BLR}$ is in units of light-days. 
\begin{table}
\begin{center}
\begin{tabular}{lccc}
\hline
Method&$\lambda$&b&a\\ 
\hline
WLS        &5100\AA&$0.57\pm0.02$&$-19.53\pm0.80$\\
FITxy	   &5100\AA&$0.60\pm0.02$&$-20.75\pm0.92$\\
BCES       &5100\AA&$0.61\pm0.10$&$-21.00\pm3.67$\\
\hline
 & &\\
WLS        &3000\AA&$0.50\pm0.02$&$-17.19\pm0.70$\\
FITxy	   &3000\AA&$0.53\pm0.02$&$-18.08\pm0.82$\\
BCES       &3000\AA&$0.47\pm0.05$&$-16.10\pm1.69$\\
\hline
\end{tabular}
\end{center}
\caption{Results of the regression analysis of the correlation between
BLR radius and the AGN continuum luminosity at 5100\AA\, and
3000\AA\,. The fits were performed in $\log-\log$ space and are of the
form: $\log{R_{BLR}}=b\log(\lambda L_{\lambda}) +a$. The three
different regression methods are discussed in the text.} 
\label{rltab}
\end{table}
\section{Estimating the broad-line velocity}
\label{proxy}
\begin{figure}
\centerline{\epsfig{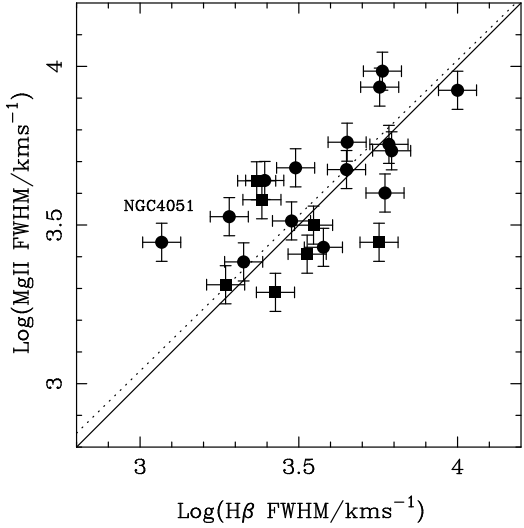}}
\caption{A Log-Log plot of \mg FWHM versus \hb FWHM for the 22 objects 
from the RM sample for which it was possible to obtain measurements of both
line-widths. The solid line is an exact 1:1 relation. The dotted line is
the BCES bisector fit, excluding NGC 4051, and has slope of
$1.02\pm0.14$. A 15\% error in the FWHM of \hb and \mg has been
assumed. Symbols are as Fig \ref{rl}.}
\label{mg2-hb_rm}
\end{figure}

After successfully calibrating the $R_{BLR}-\lambda L_{3000}$ relation, 
the second required element of the UV black-hole mass
estimator is to calibrate the \mg FWHM as a substitute for \hb. As
discussed in Section \ref{intro}, the principle reason for adopting
\mg as the UV \hb proxy is that it is a strong, fully-permitted, 
low-ionization line, and as such should provide the best UV analog of
\hb. Moreover, because the ionization potentials
of \hb and \mg are very similar, the two lines should both 
be emitted at approximately the same radius from the central ionizing source.
Fortunately, this assumption can be tested directly using the 22 
objects from the RM sample for which we
have \mg FWHM measurements. If \mg and \hb are being produced at the
same broad-line radius then, if the BLR is virialized, we should see a
1:1 relation between \mg FWHM and \hb FWHM. Fig \ref{mg2-hb_rm} 
shows Mg{\sc ii} FWHM versus H$\beta$ FWHM for the 22 objects from 
the RM sample with \mg FWHM measurements. The solid line in Fig
\ref{mg2-hb_rm} shows an exact 1:1 relation between \mg and \hb FWHM. It
can be seen that, with the exception of NGC 4051, the objects in
the RM sample are perfectly consistent with \mg and \hb tracing the
same BLR velocity. The clear outlier, NGC 4051, is a well
studied example of a narrow-line Seyfert 1 (NLS1). The \hb FWHM for
this object has been determined by numerous authors and is 
consistently found to be around 1000 kms$^{-1}$ (eg. Wandel, Peterson \& Malkan
(1999) find \hb FWHM=1170 kms$^{-1}$). However, from our line-fitting
of IUE archive spectra we find a strong broad component to
the \mg emission line with FWHM=2790$~$kms$^{-1}$. 

In fact, several
previous studies have found that the UV emission
lines of NLS1 possess strong broad components which are not seen in
their Balmer lines (eg. Rodr\'{\i}guez-Pascual,
Mas-Hesse \& Santos-Lle\'{o} 1997, Zheng et al. 1995). This suggests 
that \mg may actually provide a better estimate of the BLR velocity of
low-ionization gas in NLS1, where the Balmer lines could be biased, at 
least in part, by inclination effects. Irrespective of the reason why
NGC 4051 is an outlier in Fig \ref{mg2-hb_rm}, we can safely exclude it
from a regression fit to the \mg - \hb FWHM relation. The reason for
this is that we are principally interested in providing a UV
black-hole mass estimator for high redshift quasars, the vast majority
of which have \hb FWHM $>2000$ kms$^{-1}$. The dotted line in 
Fig \ref{mg2-hb_rm} shows the BCES bisector fit which is equivalent to
a relation of the form: 
\begin{equation}
H\beta($FWHM$) \propto MgII($FWHM$)^{1.02\pm0.14}
\end{equation}
\noindent
which is clearly consistent with a linear relation. In the light 
of this result, an 
exact 1:1 relation between the \mg FWHM and \hb FWHM
is assumed in the next section for the final calibration of the UV
black-hole mass estimator.
\section{The UV black-hole mass estimator}
\label{uv}
\begin{figure}
\centerline{\epsfig{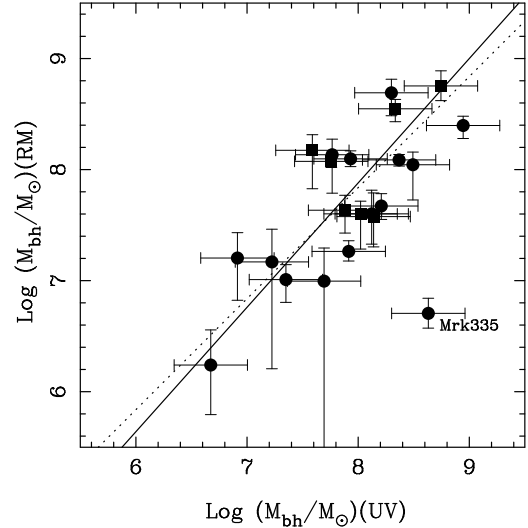}}
\caption{Full reverberation mapping black-hole mass
estimate (based on $R_{BLR}$ measurements and the rms H$\beta$ FWHM)
plotted against the UV mass estimate (based on the $R_{BLR}-\lambda
L_{3000}$ relation of Eqn 2 and the FWHM of Mg{\sc ii}) for the 
22 objects from the RM sample for which it was 
possible to obtain Mg{\sc ii} FWHM
measurements. The solid line is the BCES bisector fit, excluding
Mrk 335, which has a slope of $1.12\pm0.22$. The dotted line is the 
adopted linear relation (Eqn 6). Symbols as in Fig \ref{rl}.}
\label{test}
\end{figure}

Having determined in the previous two sections the $R_{BLR}-\lambda
L_{3000}$ relation and the 1:1 scaling between \hb FWHM and
\mg FWHM, we are now in a position to derive our UV black-hole mass estimator.
In Fig \ref{test} we show the reverberation
mapping black-hole mass estimate versus the new UV mass estimate, which uses 
the $R_{BLR}-\lambda L_{3000}$ relation to estimate
$R_{BLR}$ and the Mg{\sc ii} FWHM to trace the BLR velocity. It is
clear from Fig \ref{test} that Mrk 335 is an outlier. As with NGC 4051,
Mrk 335 is another example of a NLS1. The reason Mrk 335 is an outlier
in Fig \ref{test} is that the FWHM of the variable component of its 
\hb line determined from reverberation mapping is only 1260 kms$^{-1}$
(Wandel, Peterson \& Malkan 1999), while from our line-fitting of
IUE archive spectra we detect a strong broad component in the \mg
line with FWHM=8605 kms$^{-1}$. Our line-fitting results for this
object are in excellent agreement with those of Zheng et al. (1995)
who detected a broad \mg component of FWHM=8218 kms$^{-1}$. Again, as
mentioned in the previous section, it appears a possibility 
that for at least some 
NLS1's the width of the \hb emission line may not be a reliable tracer of
BLR velocities. Of course, it is also possible that Mrk 335 may simply
be a genuine outlier. The solid line in Fig \ref{test} is the bisector 
fit to the data,
excluding Mrk 335, and has the form:
\begin{displaymath}
\log M_{bh}(RM)=1.12(\pm0.22)\log M_{bh}(UV)
\end{displaymath}
\begin{equation}
$\hspace{3.8cm}$-1.08(\pm1.75)
\end{equation}
again, consistent with a linear relation. Consequently, 
our final calibration of the UV mass estimator has the form:
\begin{equation}
\log M_{bh}(RM)=\log M_{bh}(UV)-0.16
\end{equation}
\noindent
where the off-set of $0.16$ is adopted to ensure a mean 
$M_{bh}(RM)$:$M_{bh}(UV)$ ratio of unity. In terms of a useful formula 
the best fitting calibration of the UV black-hole mass estimator is therefore:
\begin{equation}
\frac{ M_{bh}} {\Msolar}  =3.37\left(\frac{\lambda
L_{3000}}{10^{37}{\rm W}}\right)^{0.47}\left(\frac{FWHM(MgII)}
{{\rm kms}^{-1}}\right)^{2}
\label{final}
\end{equation}
\noindent
For completeness we have also re-derived the optical (H$\beta$) virial 
black-hole mass estimator using our re-analysis of the $R_{BLR}-\lambda
L_{5100}$ relation in Section \ref{radius}. As expected, the
correlation between $M_{bh}(RM)$ and $M_{bh}(H\beta)$ is also 
consistent with being linear ($M_{bh}(RM) \propto
M_{bh}(H\beta)^{0.87\pm0.18}$) which, in the same fashion as for the
UV estimate, leads to:
\begin{equation}
\frac{M_{bh}}{\Msolar}=4.74\left(\frac{\lambda
L_{5100}}{10^{37}{\rm W}}\right)^{0.61}\left(\frac{FWHM(H\beta)}
{{\rm kms}^{-1}}\right)^{2}
\label{optical}
\end{equation}
\subsection{Accuracy of the UV estimator}
Having arrived at the final calibration of the UV black-hole mass 
estimator it is of obvious interest to assess the accuracy with which it can
reproduce the full reverberation mass estimate. Excluding Mrk 335, 
the mean difference between the reverberation and the UV estimator is :
\begin{equation}
<\log(M_{bh})(RM)-\log(M_{bh})(UV)>=0.00\pm0.40
\end{equation}
\noindent
where the quoted uncertainty is the standard deviation ($\sigma$) 
and not the standard error. Therefore, provided the RM sample 
is representative of broad-line AGN in general, we conclude that 
the UV black-hole estimator provided by Eqn$~$\ref{final} can 
reproduce the reverberation black-hole mass to within a factor 
of 2.5 $(1\sigma)$. For the same sample the uncertainty on the
optical estimator given in Eqn \ref{optical} is a factor of $2.7$.
\section{Application to quasar surveys: the LBQS and MQS}
\label{lbqs}
Having derived the desired black-hole mass estimator in the
previous section, we next assess its performance when
applied to the spectra of the LBQS and
the radio-selected MQS. If the new UV black-hole mass
estimator is to be used in the analysis of high-redshift quasars, 
it is clearly important to ensure that it does not produce
results which are biased with respect to the usual optical (H$\beta$) 
estimator. Therefore, our intention in this section is
to simply test how the new UV mass
estimator (Eqn \ref{final}) compares to the usual optical mass estimator
(Eqn \ref{optical}) when applied to those objects from the LBQS and MQS
for which both the appropriate UV ($\lambda L_{3000}$, \mg FWHM) and
optical ($\lambda L_{5100}$, \hb FWHM) data are available. 

Due to the fact that the LBQS is optically 
selected, while the MQS is a complete
radio-selected quasar sample, applying the UV black-hole mass
estimator to both samples allows us to ensure that Eqn \ref{final} is
equally applicable to both radio-quiet and radio-loud quasars.
 
\subsection{The Large Bright Quasar Survey}
The LBQS consists of 1058 optically-selected quasars in the redshift range
$0.2$~$<z<$~$3.3$ (see Hewett, Foltz \& Chaffee (1995) for a full description).
A comprehensive study of the optical continuum and 
emission line properties of 992 quasars from the LBQS has recently been 
published by Forster et al. (2001). In this section we use data from
Forster et al. for a 99-object sub-sample with reliable measurements 
of both \mg and \hb FWHM and continuum fluxes at 3000\AA\, and 
5100\AA\ . This sub-sample comprises 68\% of the 145 
LBQS objects in the redshift range $0.20<z<0.66$ within which 
both \mg and \hb fall on the LBQS spectra.

\subsection{The Molonglo quasar sample}
The Molonglo quasar sample is a complete, radio-selected, sample consisting 
of 111 quasars in the redshift range $0.1$~$<$~$z$~$<2.9$ with a 408 MHz flux
density greater than 0.95 Jy (Kapahi et al. 1998). Optical spectra for
79 MQS quasars were published by Baker et al. (1999), 
together with a preliminary analysis of their emission line
properties. We are currently engaged in an automated study of the
optical and UV emission lines from the MQS optical spectra 
(kindly provided by J. Baker), from which it has been possible to reliably
measure the \mg and \hb FWHM for 29/38 objects where 
the MQS spectra include both emission lines. Unlike the LBQS, Baker
(1997) has shown that many MQS objects display substantial reddening
in their optical spectra. As a result, the continuum luminosities
of the reddest MQS objects were de-reddened using the SMC
extinction curve of Pei (1992), assuming an intrinsic continuum slope
of $\alpha=0.5$ ($f\nu \propto \nu^{-\alpha}$) which was found by
Baker (1997) to be typical of the least reddened MQS objects.

\subsection{Optical versus UV black-hole mass estimators}
In Fig \ref{mg2-hb} we show the optical black-hole mass estimator plotted
against the new UV black-hole mass estimator for a combined sample of
150 objects, comprising 99 from the LBQS, 29 from the MQS and 
the 22 objects of the RM sample. Also shown is the BCES 
bisector fit to the LBQS and MQS objects which has the form: 
\begin{figure}
\centerline{\epsfig{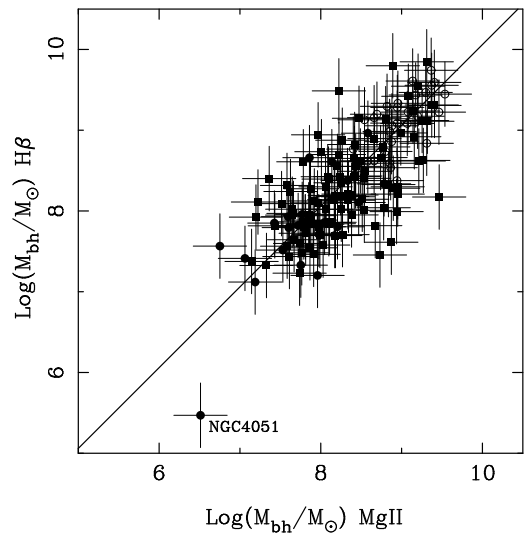}}
\caption{The optical (H$\beta$) versus UV (Mg{\sc ii}) virial black-hole
estimators for 150 objects from the RM (filled circles), LBQS (filled
squares) and MQS (open circles) samples. The solid line is the
BCES bisector fit to the 128 objects from the MQS and LBQS samples and
has a slope of $1.00\pm0.08$. The outlier NGC4051 has been highlighted
(see Section 5).}
\label{mg2-hb}
\end{figure}
\begin{displaymath}
\log M_{bh}(H\beta)=1.00(\pm0.08)\log M_{bh}(MgII)
\end{displaymath}
\begin{equation}
$\hspace{3.8cm}$+0.06(\pm0.67)
\end{equation}
\noindent
which, as expected, is perfectly consistent with a linear relation. In Fig
\ref{ratio} we show a histogram of 
$\log M_{bh}(MgII)-\log M_{bh}(H\beta)$
for the 128 objects from the LBQS and MQS. The solid line shows the
best-fitting gaussian which has $\sigma=0.41$. These results, in
combination with those of Section \ref{uv}, lead us to conclude that 
compared to the traditional optical black-hole mass estimator, the 
new UV estimator provides results which are unbiased and of equal
accuracy.
\begin{figure}
\centerline{\epsfig{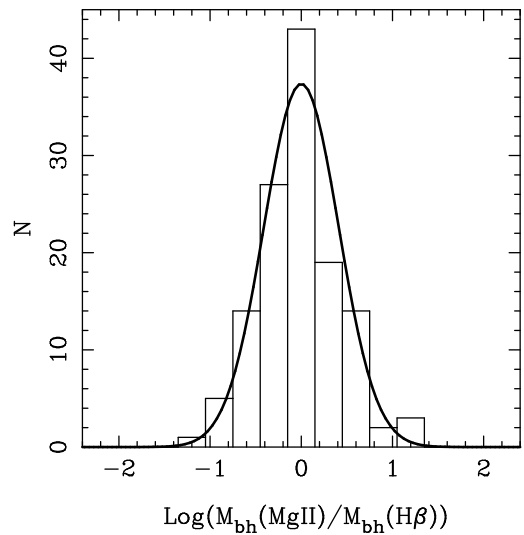}}
\caption{Histogram of $\log M_{bh}(MgII)-\log M_{bh}(H\beta)$ for 
the 128 objects
from the LBQS and MQS shown in Fig \ref{mg2-hb}. Also shown is the 
best-fitting gaussian which has $\sigma=0.41$.}
\label{ratio}
\end{figure}

\subsection{Black-hole mass and the radio-loudness dichotomy}
Although a full analysis of the black-hole masses of the LBQS and 
MQS samples is beyond the scope of this paper, we will briefly comment 
here on one obvious feature of Fig~4. It is immediately apparent 
from Fig \ref{mg2-hb} that the MQS quasars are confined to the highest
black-hole masses. In fact, 27/29 of the MQS quasars have black-hole
masses $\geq 10^{8.5}\Msolar$, adopting either the UV or optical 
mass estimator. In contrast, only 17\% of the 99 LBQS quasars 
have UV and optical black-hole mass estimates which are 
both $\geq10^{8.5}\Msolar$. Naively this would appear to suggest that
there is a clear division in black-hole mass between the
radio-selected 
MQS and the largely radio-quiet LBQS. However, because 
the wavelength coverage of the LBQS spectra only extends to
$\sim7500$\AA, compared to $\sim 10000$\AA\, for those of the MQS,
the need to include both \mg and \hb on the spectra introduces 
a considerable redshift bias into the two sub-samples under
investigation here. Due to the differences in wavelength coverage the
mean redshift of the LBQS sub-sample is $0.33\pm0.01$, while that of 
the MQS is $0.63\pm0.03$. Consequently, the members of the MQS
sub-sample are on average more optically luminous than those of the
LBQS sub-sample, with an average luminosity of 
$\log(\lambda L_{5100}/{\rm W})=38.14\pm0.08$
compared to $\log(\lambda L_{5100}/{\rm W})=37.64\pm0.07$. Therefore,
with reference to the $R_{BLR}-\lambda L_{\lambda}$ relation, we would 
expect that, simply due to this luminosity bias, the average 
black-hole mass of the MQS
sub-sample would be a factor of $\sim2$ larger than that of the
LBQS. However, in reality the average black-hole mass of the MQS 
sub-sample is $10^{9.10\pm0.07}\Msolar$ whereas that of the LBQS
sub-sample is more than a factor of 6 lower at 
$10^{8.27\pm0.06}\Msolar$\footnote{These figures are for the optical
(H$\beta$) mass estimator although, as expected from Fig \ref{mg2-hb},
the black-hole mass difference is comparable if the 
UV mass estimator is adopted ($\Delta M_{bh}=0.71$ dex)}. 

The origin of the additional factor of three difference in the average
black-hole masses is a highly significant difference in the
distributions of \hb and \mg FWHM. For example, the average
\mg FWHM of the MQS sub-sample is a factor of $1.56\pm0.11$ greater
than the average \mg FWHM of the LBQS sub-sample. The equivalent
factor for \hb FWHM is $1.68\pm0.14$. The application of a
Kolmogorov-Smirnov test returns a probability of $p=1.33\times 10^{-6}$
that the two \hb FWHM distributions are drawn from the same parent
distribution ($p=4.1\times 10^{-9}$ for the \mg FWHM
distributions). 

It is important to note that the differences in the FWHM
distributions of the two samples do not appear to be related to 
the previously mentioned redshift/luminosity bias. An application of
the partial spearman rank correlation test (Macklin
1982) reveals no significant correlations between \hb FWHM, redshift 
and $\lambda L_{5100}$ or \mg FWHM, redshift and $\lambda
L_{3000}$. In fact, these results are in good agreement with previous
studies of the black-hole masses of radio-quiet and radio-loud quasars
which were not subject to a redshift/luminosity bias. In their study
of a sample of radio-loud and radio-quiet quasars with a matched
redshift-luminosity distribution at $z\simlt 0.5$, 
McLure \& Dunlop (2002) found the average black-hole mass of the 
radio-loud quasars to be a factor of $\simeq2$ larger than their
radio-quiet counterparts, albeit with a large overlap. Likewise, Laor
(2000) found a clean separation in black-hole mass within the PG
quasar sample, with the radio-loud quasars being confined to
$M_{bh} \simgt 10^{9}\Msolar$. In contrast, from their study
of the First Bright Quasar Survey (FBQS), Lacy et al. (2001) 
concluded that the apparent gap in the quasar 
black-hole mass -- radio-power plane is in fact filled by previously
undetected radio-intermediate quasars. 

In either case, it is 
clear that genuinely powerful, low-frequency selected, radio-loud  quasars 
harbour central black-hole masses drawn from the 
extreme end of the AGN black-hole mass function. At present it is 
not clear whether the low frequency of radio-quiet quasars 
with $M_{bh}>10^{8.5}\Msolar$ is real or, alternatively, is due to 
an inclination selection effect whereby the line-widths of optically 
selected quasars are biased to low values by a preference for 
selecting objects close to the line of sight. A more detailed analysis 
of the black-hole masses of high redshift radio-loud and radio-quiet 
quasars, the correlation between black-hole mass and radio power and 
the evidence for the role of inclination effects will be presented 
in a series of forthcoming papers (McLure et al. 2002, in prep).

Finally, we
note that our finding that the FWHM of \hb and \mg in the powerful 
radio-selected MQS objects are exclusively restricted to $\simgt 4000$
kms$^{-1}$ is of interest in the context of recent work on the
location of radio-loud and radio-quiet AGN in so-called eigenvector~1 
space (Marziani et al. 2001). In their study Marziani et al. concluded 
that powerful radio-loud quasars appear to be restricted to 
\hb FWHM $>4000$ kms$^{-1}$, in excellent agreement with our analysis of 
the MQS spectra.

\section{Conclusions}
\label{conc}
A new technique for estimating the central black-hole masses of 
high-redshift quasars using the \mg FWHM and 3000\AA\, quasar continuum 
luminosity has been presented. The new technique has been calibrated
using a sample of 34 low redshift AGN with black-hole mass measurements 
based on long-term reverberation mapping experiments. The reliability of 
the new technique, with respect to the established optical virial black-hole
 mass estimator, has been tested using published data for the LBQS
together with the results of a new analysis of the emission line
properties of the MQS. The main conclusions of this study can be summarized as follows:
\begin{itemize}

\item{The correlation between $R_{BLR}$ and monochromatic
3000\AA\, continuum luminosity is found to display less scatter than
that between $R_{BLR}$ and 5100\AA\, monochromatic continuum luminosity.}
\item{The correlation between $R_{BLR}$ and 3000\AA\, continuum
luminosity is found to be consistent with a relation of the 
form $R_{BLR}\propto \lambda L_{\lambda}^{0.5}$, as expected for a 
constant ionization parameter in all AGN, irrespective of luminosity}
\item{The FWHM of \mg is found to be an 
effective substitute for the FWHM of \hb. The relationship between 
the two FWHM is found to be perfectly consistent with an exact 1:1 scaling.}
\item{Combining the $R_{BLR}-\lambda L_{3000}^{0.47}$ relation with the
FWHM of \mg produces a virial black-hole mass estimator based on
rest-frame UV observables which is capable of reproducing black-hole
masses determined from reverberation mapping to 
within a factor of 2.5 ($1\sigma$)}
\item{An application to objects from the LBQS and MQS 
demonstrates that the new UV black-hole mass estimator produces results
which are unbiased, and of equal accuracy to the established 
optical (H$\beta$) black-hole mass estimator.}
\item{We therefore conclude that the new UV black-hole mass estimator
is ideal for determining quasar black-hole masses in the redshift 
range $0.25<z<2.5$ via optical spectroscopy alone.}
\end{itemize}
\section{acknowledgments}
The authors are happy to acknowledge Jo Baker for providing the optical
spectra of the MQS. Dan Maoz and James Dunlop are acknowledged for 
useful discussions. RJM acknowledges the
award of a PPARC postdoctoral fellowship. MJJ acknowledges the support
of the European Community Research and 
Training Network `The Physics of the Intergalactic Medium'. This
research is in part based on INES data from the IUE satellite.This
research has made use of 
the NASA/IPAC Extragalactic Database (NED)
which is operated by the Jet Propulsion Laboratory, California Institute
of Technology, under contract with the National Aeronautics and Space
Administration. 


\begin{thebibliography}{46}
\bibitem{1} Akritas M.G., Bershady M.A., 1996, ApJ 470, 706
\bibitem{2} Baker J.C., 1997, MNRAS, 286, 23
\bibitem{3} Baker J.C., Hunstead R.W., Kapahi V.K., Subrahmanya C.R.,
1999, ApJS, 122, 29
\bibitem{4} Barthel P.D., 1989, ApJ, 336, 606
\bibitem{5} Boroson T.A., Green R.F., 1992, ApJS, 80, 109
\bibitem{6} Brotherton M.S., ApJS, 1996, 102, 1
\bibitem{7} Corbin M.R., Boroson T.A., 1996, ApJS, 107, 69 
\bibitem{8} Hewett P.C., Foltz C.B., Chaffee F.H., 1995, AJ, 109, 1498
\bibitem{9} Dunlop J.S., McLure R.J., Kukula M.J., Baum S.A., 
O'Dea C.P., Hughes D.H., 2002, MNRAS, submitted, astro-ph/0108397
\bibitem{10} Forster K., Green P.J, Aldcroft T.L., Vestergaard M., Foltz C.B.,
Hewett P.C., 2001, ApJS, 134, 35
\bibitem{11} Ferrarese L., Pogge R.W., Peterson B.M., Merritt D., 
Wandel A., Joseph C.L., 2001, ApJ, 555, L79
\bibitem{12} Gebhardt K., et al., 2000, ApJ, 543, L5
\bibitem{13} Kapahi V.K., Athreya R.M., Subrahmanya C.R., Baker J.C.,
Hunstead R.W., McCarthy P.J., van Breugel W., 1998, ApJS, 118, 327
\bibitem{14} Kaspi S., Smith P.S., Netzer H., Maoz D., Jannuzi B.T., Giveon U.,
2000, ApJ, 533, 631
\bibitem{15} Krolik J.H., 2001, ApJ, 551, 72 
\bibitem{16} Lacy M., Laurent-Muehleisen S.A., Ridgway S.E., 
Becker R.H., White R.L., 2001, ApJ, 551, L17
\bibitem{50} Laor A., 2000, ApJ, 543, L111
\bibitem{17} Laor A., 2001, ApJ, 553, 677
\bibitem{18} McLure R.J., Dunlop J.S., 2001, MNRAS, 327, 199
\bibitem{19} McLure R.J., Dunlop J.S., 2002, MNRAS, 331, 795 
\bibitem{20} Macklin J.T., 1982, MNRAS, 199, 1119
\bibitem{21} Marziani P., Sulentic J.W., Dultzin-Hacyan D., Calvani
M., Moles M., 1996, ApJS, 104, 37
\bibitem{22} Marziani P., Sulentic J.W., Zwitter T., Dultzin-Hacyan
D., Calvani M., 2001, ApJ, 558, 553
\bibitem{23} Neugebauer G., Green R.F., Matthews K., Schmidt M.,
Soifer B.T., Bennett J., 1987, ApJS, 63, 615
\bibitem{24} Onken C.A., Peterson B.M., 2002, ApJ, in press, astro-ph/0202382
\bibitem{25} Pei Y.C., 1992, ApJ, 395, 130
\bibitem{26} Peterson B.M., Wandel A., 2000, ApJ, 540, L13
\bibitem{27} Peterson B.M., et al., 2000, ApJ, 542, 161
\bibitem{28} Press W.H., Teukolsky S.A., Vetterling W.T., Flannery B.P., 1992,
Numerical Recipes, Cambridge University Press
\bibitem{29} Rodr\'{\i}guez-Pascual P., Mas-Hesse J.M., Santos-Lle\'{o} M.,
1997, A\&A, 327, 72
\bibitem{30} Vestergaard M., Wilkes B.J., Barthel P.D., 2000, ApJ, L103 
\bibitem{31} Vestergaard M., 2002, ApJ, 571, 733
\bibitem{32} Wandel A., Peterson B.M., Malkan M.A., 1999, ApJ, 526, 579
\bibitem{33} Wills B.J., Browne I.W.A., 1986, ApJ, 302, 56 
\bibitem{34} Zheng W., et al., 1995, ApJ, 444, 632
\end{thebibliography}
\end{document}